\documentclass[twocolumn]{aastex631}
\usepackage{xcolor, mathtools, graphicx, dcolumn, bm, hyperref, tikz, etoolbox}

\definecolor{myRed}{rgb}{.84,0,.08}
\definecolor{myGreen}{rgb}{.14,.54,.24}
\definecolor{myBlue}{rgb}{0,.25,.87}
\newcommand{\bs}[1]{\boldsymbol{#1}}
\def\be{\begin{equation}}
\def\ee{\end{equation}}
\def\ba{\begin{eqnarray}}
\def\ea{\end{eqnarray}}
\def\pdt{\partial_t}
\def\div{\nabla \cdot}
\def\rot{\nabla \times}
\def\Di{\mathcal{D}_\mathrm{I}}
\def\Dis{\mathcal{D}_\mathrm{I}^\sigma}
\def\meanDi{\left\langle \mathcal{D}_\mathrm{I} \right\rangle}
\def\meanDis{\left\langle \mathcal{D}_\mathrm{I}^\sigma \right\rangle}

\begin{document}


\title{Energy transfer, discontinuities and heating in the inner solar wind measured with a weak and local formulation of the Politano-Pouquet law}


\author{V. David}
\affiliation{Laboratoire de Physique des Plasmas (LPP), Université Paris-Saclay, CNRS, École Polytechnique, Institut Polytechnique de Paris, Sorbonne Université, Observatoire de Paris, 91120 Palaiseau, France}%

\author{S. Galtier}
\affiliation{Laboratoire de Physique des Plasmas (LPP), Université Paris-Saclay, CNRS, École Polytechnique, Institut Polytechnique de Paris, Sorbonne Université, Observatoire de Paris, 91120 Palaiseau, France}%
\affiliation{Institut universitaire de France}

\author{F. Sahraoui}
\affiliation{Laboratoire de Physique des Plasmas (LPP), Université Paris-Saclay, CNRS, École Polytechnique, Institut Polytechnique de Paris, Sorbonne Université, Observatoire de Paris, 91120 Palaiseau, France}%

\author{L.\,Z. Hadid}
\affiliation{Laboratoire de Physique des Plasmas (LPP), Université Paris-Saclay, CNRS, École Polytechnique, Institut Polytechnique de Paris, Sorbonne Université, Observatoire de Paris, 91120 Palaiseau, France}%


\begin{abstract}

The solar wind is a highly turbulent plasma for which the mean rate of energy transfer $\varepsilon$ has been measured for a long time using the Politano-Pouquet (PP98) exact law. However, this law assumes statistical homogeneity that can be violated by the presence of discontinuities. Here, we introduce a new method based on the inertial dissipation $\Dis$ whose analytical form is derived from incompressible magnetohydrodynamics (MHD); it can be considered as a weak and {\it local} (in space) formulation of the PP98 law whose expression is recovered after integration is space. We used $\Dis$ to estimate the local energy transfer rate  from the \textit{THEMIS-B} and \textit{Parker Solar Probe} (PSP) data taken in the solar wind at different heliospheric distances. Our study reveals that discontinuities near the Sun lead to a strong energy transfer that affects a wide range of scales $\sigma$. We also observe that switchbacks seem to be characterized by a singular behavior with an energy transfer varying as $\sigma^{-3/4}$, which slightly differs from classical discontinuities characterized by a $\sigma^{-1}$ scaling. A comparison between the measurements of $\varepsilon$ and $\Dis$ shows that in general the latter is significantly larger than the former.

\end{abstract}


\section{Introduction}
\label{sec:intro}
For several decades, the solar wind --  a collisionless plasma -- has been the subject of an apparent paradox. The measurements made by Voyager 1 \& 2 revealed that the average (proton) temperature of the solar wind decreases as $\sim r^{-0.5}$ over $1$--$20$ Astronomical Units (AU), with $r$ the radial distance from the Sun \citep{Gazis1982,Marsch1982,Richardson1995,Matthaeus99}. However, for a radially-expanding, adiabatically cooling plasma, one would expect a temperature variation as $r^{-4/3}$, which is significantly steeper than the observed law. This paradox can be solved if an efficient local heating source exists, which must be collisionless in nature (note, however, that the adiabatic model can be questioned since it derives from a fluid approximation, which implicitly assumes the existence of collisions).

In the near outer heliosphere ($r > 2$\,AU), large-scale shocks (or stream shear as a source of turbulence) at the interface between high and low speed streams were quickly suspected as a major source of heating \citep{Gazis1982,Burlaga1987,DG21}. In the far outer solar wind ($r>20$\,AU) where the temperature increases slightly \citep{Matthaeus99,Elliott2019}, pick up ions are considered as a main source of heating \citep{Gazis1994,Pine2020}. These are originally neutrals from the interstellar medium that are transformed into ions by charge exchange with solar wind protons, and are eventually picked up by the interplanetary magnetic field. In this context, several (phenomenological) turbulence transport model equations have been successfully used to study the solar wind heating  \citep{Zank1996,Zank2018}.

In the inner heliosphere ($r \le 1$\,AU), the situation is different because turbulent fluctuations are dominant. (By turbulent fluctuations, we mean a medium not dominated by large scale structures like the interplanetary shocks observed at 5\,AU: in this case, fluctuations are also detected but only as a small-scale modification of the shocks.) Therefore, studies focus on the turbulent cascade which is seen as an efficient mechanism to bring energy from large magnetohydrodynamic (MHD) scales to small kinetic (sub-MHD) ones \citep{Sahraoui20}. {\it In-situ} measurements of $\varepsilon$, the mean rate of energy transfer at MHD scales, provides an estimate of the heating rate by assuming complete conversion from the former to the latter. While those estimates cannot inform us about the precise kinetic mechanism responsible for energy dissipation, recent progress using Landau-fluid simulations showed the ability of the exact laws to estimate the amount of dissipation due to Landau damping \citep{Ferrand2021b}. 

In practice, $\varepsilon$ can be estimated from exact laws. First developed in incompressible hydrodynamics \citep{K41,Batchelor53,Antonia97}, the exact laws have been derived for many physical systems where turbulence is encountered. This includes isothermal compressible hydrodynamics \citep{G11}, a model often used to simulate supersonic interstellar turbulence \citep{Kritsuk2007,Federrath2010,Ferrand2020}. For the solar wind, the simplest exact law is that derived from incompressible MHD \citep{PP98a}. Its use led to the first estimate of turbulent heating in the solar wind \citep{SorrisoValvo07,MacBride08,Marino08,Stawarz09,Stawarz10,Osman2011}. Later, several generalized exact laws were derived to account for compressible MHD \citep{BG13, Andres17,Simon2021}, Hall-MHD \citep{G08a,BG17b,Andres18, Hellinger2018,Ferrand2021} and even gravito-turbulence \citep{BK17a,BK17b}. 
With these new laws, it was possible to obtain better estimates of $\varepsilon$ in the solar wind and planetary plasma environments that incorporate density fluctuations and sub-ion scale effects \citep{Banerjee2016,Hadid17,Andres2019,Bandyopadhyay2020b, Andres2021}. 

Exact laws are based on the zeroth law of turbulence (unproved in general) which says that in a turbulence experiment, everything else being fixed, if the energy dissipation ends to zero, the mean rate of energy dissipation tends to a non-zero limit, which is $\varepsilon$ \citep{Frisch95}. This law has led to an interesting mathematical development around the concept of weak solutions in Euler's equation, useful when the velocity becomes non-regular \citep{Leray34}. In particular, the non-regularity of the field can lead in principle to energy dissipation without the assistance of viscosity \citep{Onsager49}. This new form of dissipation has been called inertial dissipation (noted hereafter $\Di$) as opposed to viscous dissipation. The mathematical expression of $\Di$ for the Euler equation \citep{Duchon} has a striking similarity with Kolmogorov's law \citep{Antonia97}. Unlike the exact law, the expression of $\Di$ does not involve an ensemble average and, therefore, can be used at any point in a turbulent fluid to evaluate the local (in space) dissipation \citep{Saw2016}. This work on incompressible hydrodynamics has recently been generalized to 3D incompressible (Hall) MHD \citep{G18} and to a low dimensional MHD system \citep{Yanase97} that has been used to estimate the inertial dissipation produced by collisionles shocks in the outer heliosphere \citep{DG21}. Like with Burgers' equation \citep{EyinkNotes,Dubrulle19}, with the low dimensional MHD model the zeroth law of turbulence can be proved with, on average, $\langle \Di \rangle = \varepsilon$. 

The structure of the paper is as follows. Section \ref{sec:theoreticalModel} is devoted to theoretical framework (incompressible MHD, exact law, inertial dissipation). Section \ref{sec:methods} presents the selection of data (THEMIS-B, PSP) and their processing; various situations are considered (slow and fast winds, discontinuities). The results of our analysis are presented in Section \ref{sec:results} with in particular the measurements of $\varepsilon$ and $\Di$. A conclusion is finally given in Section \ref{sec:conclusion}. 


\section{MHD theory}
\label{sec:theoreticalModel}
\subsection{Four-thirds exact law}

We briefly recall the four-thirds exact law for incompressible MHD derived by \cite{PP98a}, which we will hereafter call the PP98 law.
Let $\bs{u}$ be the fluid velocity, $\bs{b} \equiv \bs{B}/\sqrt{\mu_0 \rho_0}$ the magnetic field normalized to a velocity with $\rho_0$ the mean plasma density and $\mu_0$ the vacuum permeability, $P_*=P+b^2/2$ the sum of the thermal and magnetic pressures, $\nu$ the kinematic viscosity and $\eta$ the magnetic diffusivity. Then, the incompressible MHD equations read \citep{G16}
\ba
\label{eq:dudt}
\pdt \bs{u} + \bs{u} \cdot \nabla \bs{u} &=& - \nabla P_* + \bs{b} \cdot \nabla \bs{b} + \nu \nabla^2 \bs{u}, \\
\label{eq:dbdt}
\pdt \bs{b} +\bs{u} \cdot \nabla \bs{b} &=& \bs{b} \cdot \nabla \bs{u} + \eta \nabla^2 \bs{b},
\ea
where $\bs{u}$ and $\bs{b}$ are zero-divergence fields. To derive these equations, the following Ohm’s law is used
\be \label{eq:ohmLaw}
\bs{e} = \eta \bs{j} - \bs{u} \times \bs{b},
\ee
where $\bs{e}$ is the normalized electric field and $\bs{j} = \rot \bs{b}$ the normalized electric current density. To obtain the PP98 law, we assume a large-scale stationary forcing and an asymptotically large (magnetic and kinetic) Reynolds numbers.
After a standard calculation, one obtains a primitive form of the PP98 exact law
\be \label{eq:epsilon1D}
- 4 \varepsilon  = \nabla_\ell \cdot \left\langle \left( \left\vert \delta \bs{u} \right\vert^2 + \left\vert \delta \bs{b} \right\vert^2 \right) \delta \bs{u} - 2 \left( \delta \bs{u} \cdot \delta \bs{b} \right) \delta \bs{b} \right\rangle,
\ee
where $\left\langle \cdot \right\rangle$ is the ensemble average. For any variable $g$, $\delta g \equiv g \left( \bs{x} + \bs{\ell} \right) - g \left( \bs{x} \right)$, with $\bs{\ell}$ the vector increment. In this expression, $\varepsilon$ is the mean rate of energy transfer/dissipation/forcing, the equivalence between the three definitions being due to the stationarity assumption. 

The previous expression can be reduced to the PP98 law when the statistical isotropy is further assumed
\be \label{eq:epsilon}
- \frac{4}{3} \varepsilon \ell = \left\langle \left( \left\vert \delta \bs{u} \right\vert^2 + \left\vert \delta \bs{b} \right\vert^2 \right) \delta u_\ell - 2 \left( \delta \bs{u} \cdot \delta \bs{b} \right) \delta b_\ell \right\rangle.
\ee
Here, the index $\ell$ refers to a projection along the longitudinal direction given by the vector $\bs{\ell}$, with $\ell$ its norm. The PP98 exact law is valid in the inertial range of incompressible MHD turbulence. A basic assumption made to use the law (\ref{eq:epsilon}) is that the fields are regular.
In simple terms, a field is said to be regular if all the classical tools of analysis (such as  derivative calculations) can be applied. In case of non-regular fields (e.g. a discontinuity), a weak formulation must be introduced. 

\subsection{Weak formulation}
\label{section22}
\begin{figure}[t!]
    \centering
    \includegraphics{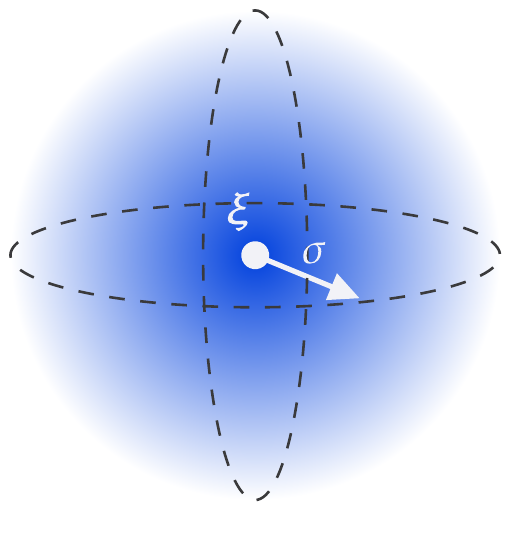}
    \caption{Scheme of the filtering process. The color reflects the intensity of the smoothing. See equation (\ref{eq:coarseGrained}) for the definitions of $\sigma$ and $\xi$.}
    \label{fig:filteringScheme}
\end{figure}

The weak formalism is based on smoothing of a field with some kernel $\varphi \in \mathbb{C}^{\infty}$ with compact support on $\mathbb{R}^3$, even, non-negative and with integral $1$. To formalize the notion of scale, we define a test function $\varphi^\sigma$ such that $\varphi^\sigma (\bs{\xi}) \equiv \sigma^{-3} \varphi(\bs{\xi}/\sigma)$. The regularized fields at scale $\sigma$ are defined by taking the convolution product of the fields with $\varphi^\sigma$ (for simplicity, the time dependence is omitted) 
\be \label{eq:coarseGrained}
\bs{u}^\sigma (\bs{x}) \equiv \ \varphi^\sigma * \bs{u} = \int_{\mathbb{R}^3} \varphi^\sigma (\bs{\xi}) \bs{u}(\bs{x}+\bs{\xi}) \mathrm{d}\bs{\xi},
\ee
which tends to $\bs{u} (\bs{x})$ when $\sigma\to 0$.
The other regularized quantities are defined in the same way. Note that this filtering process consists in smoothing the fields in a space defined by a sphere of radius $\sigma$ centered at the point $\bs{\xi}$ (see Figure\,\ref{fig:filteringScheme}). Under these considerations, the kinetic energy reads
\ba
E_u^\sigma(\bs{x}) &\equiv& \frac{1}{2} u_i u_i^\sigma = \frac{1}{2} \int_{\mathbb{R}^3} \varphi^\sigma (\bs{\xi}) u_i(\bs{x}) u_i(\bs{x}+\bs{\xi}) \mathrm{d}\bs{\xi}, 
\ea
where the Einstein summation convention is used (the generalization to the magnetic energy is straightforward). The previous expression can also be interpreted as the local equivalent of a correlation function where the ensemble average is replaced by a local average over scale.

With the above definitions and using a point-splitting regularization, one can derive the following weak formulation (valid for individual realizations) of the local energy conservation at position $\bs{x}$ \citep{G18}
\be
\pdt E^\sigma (\bs{x}) + \div \bs{\Pi}^\sigma (\bs{x}) = - \mathcal{D}_{\nu,\eta}^\sigma (\bs{x}) -\Di^\sigma (\bs{x}) , \label{locenergy}
\ee
with $E^\sigma = E_u^\sigma + E_b^\sigma$ the total energy. $\bs{\Pi}^\sigma$ is the spatial flux whose heavy form is not given explicitly here; this is a purely local term that describes how energy is transported across the flow, and it vanishes after integration over space with the appropriate boundary conditions. We also have the energy dissipation by viscous and resistive effects (that includes the vorticity $\bs{\omega} = \rot \bs{u}$)
\be
\mathcal{D}_{\nu,\eta}^\sigma (\bs{x})  = \nu \bs{\omega} \cdot \bs{\omega}^\sigma + \eta \bs{j} \cdot \bs{j}^\sigma,
\ee
and the inertial (also called anomalous or defect \citep{Eyink2003}) dissipation
\be
\Dis (\bs{x}) = \frac{1}{4} \int_{\mathbb{R}^3} \nabla \varphi^\sigma \left(\bs{\xi}\right) \cdot \bs{Y}(\bs{x},\bs{\xi}) \mathrm{d} \bs{\xi}, \label{eq:defDi}
\ee
where the third-order mixed structure function reads
\be
\bs{Y}(\bs{x},\bs{\xi}) = \left( \left\vert \delta \bs{u} \right\vert^2 + \left\vert \delta \bs{b} \right\vert^2 \right) \delta \bs{u} - 2 \left( \delta \bs{u} \cdot \delta \bs{b} \right) \delta \bs{b},
\ee
with $\delta g \equiv g \left( \bs{x} + \bs{\xi} \right) - g \left( \bs{x} \right)$. 
Expression (\ref{locenergy}) must be seen as a generalization of the PP98 law (or more precisely of the K\'arm\'an-Howarth MHD equation \citep{PP98a}) that we can recover for regular fields and homogeneous turbulence (see below). 
Note that in the limit $\sigma \to 0$, the two dissipative terms are mutually exclusive: the presence of any viscosity/resistivity should prevent the formation of singularities.
Thus, in this limit, only one of them can appear in the equation.
Another physical relevance of the weak formulation is revealed when performing an integration over space. The absence of an energy source at the boundary is formally equivalent to assuming periodicity (or homogeneity); therefore, the notation $\langle \cdot \rangle$ will be used for integration in space. We find
\be \label{eq:energyConservation}
\pdt \langle E^\sigma \rangle = - \langle \mathcal{D}_{\nu,\eta}^\sigma \rangle - \langle \Dis \rangle ,
\ee
with 
\be
\langle \Dis \rangle = \frac{1}{4} \int_{\mathbb{R}^3} \nabla \varphi^\sigma \left(\bs{\xi}\right) \cdot \langle \bs{Y}(\bs{x},\bs{\xi}) \rangle \mathrm{d} \bs{\xi} .\label{eq:defDi2}
\ee
In the small scale limit, we find for a viscous/resistive flow
\be
\lim_{\sigma \to 0} \langle \mathcal{D}_{\nu,\eta}^\sigma \rangle \equiv \langle \mathcal{D}_{\nu,\eta} \rangle = \varepsilon .
\ee
Therefore, $\mathcal{D}_{\nu,\eta}^\sigma$ can be used to trace, locally and across scales, the rate of viscous/resistive energy dissipation \citep{Kuzzay2019}. 
On the other hand, expression (\ref{eq:defDi2}) has a strong similarity with the RHS term of the exact law (\ref{eq:epsilon1D}), especially if one performs an integration by part, assuming the fields to be regular, and takes the small scale limit
\ba\label{eq15}
\Di (\bs{x}) &\equiv& \lim_{\sigma \to 0} \Dis (\bs{x}) \nonumber \\
&=& - \lim_{\sigma \to 0} \frac{1}{4} \int_{\mathbb{R}^3} \varphi^\sigma \left(\bs{\xi}\right) \nabla \cdot \bs{Y}(\bs{x},\bs{\xi}) \mathrm{d} \bs{\xi}. 
\ea
This relation connects directly $\Di$ to the PP98 law, which leads to the remarkable equality $\langle \Di \rangle = \varepsilon$ (see Appendix A).
Therefore, $\Dis$ can be used to trace, locally and across scales, the rate of energy transfer. 

Other interpretations can be made based on relation (\ref{eq15}). In presence of finite viscosity and resistivity, the fields are regular and thus satisfy $\lim_{\bs{\xi} \to \bs{0}^+} \delta \bs{u} = \lim_{\bs{\xi} \to \bs{0}^+} \delta \bs{b} = \bs{0}$, which leads to $\Di=0$; this is the classical situation. On the contrary, if $\nu=\eta=0$, the fields are non-regular and $\Di$ can have a contribution. This contribution is however not systematic because the fields must satisfy the H\"older condition \citep{Onsager49}. Using a scaling analysis (at a fixed position $\bs{x}$), we can make three theoretical predictions of practical importance:

\begin{enumerate}
    \item In the inertial range where the fields correspond to turbulent fluctuations that obey the PP98 law in the inertial range, we have $\delta u^3 \sim \delta b^3 \sim \sigma$ and thus $\Dis (\bs{x}) \sim \sigma^0$.
     \item At small scales where viscous/resistive effects dominate, a Taylor expansion gives $\delta u \sim \delta b \sim \sigma$ and thus $\Dis (\bs{x}) \sim \sigma^2$.
    \item However, when the fields are non-regular and act like discontinuities, the increments correspond to jumps $\delta u \sim \Delta_u$, $\delta b \sim \Delta_b$, and thus $\Dis (\bs{x}) \sim \sigma^{-1}$.
\end{enumerate}

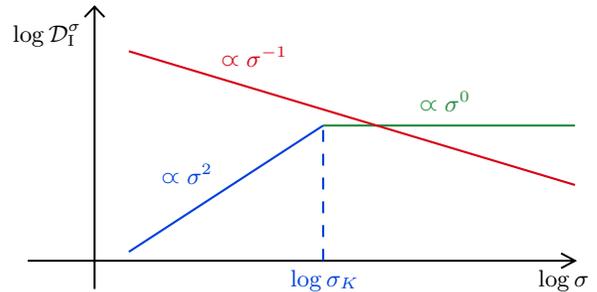
\begin{figure}[h!]
    \centering
    \tikzset{every picture/.style={line width=0.75pt}} 

\begin{tikzpicture}[x=0.75pt,y=0.75pt,yscale=-0.75,xscale=1]

\draw  (84,236) -- (360,236)(117.7,65) -- (117.7,255) (353,231) -- (360,236) -- (353,241) (112.7,72) -- (117.7,65) -- (122.7,72)  ;

\draw [color={rgb, 255:red, 36; green, 138; blue, 61 }  ,draw opacity=1  , thick]   (233,145) -- (360,145) ;
\draw [color={rgb, 255:red, 0; green, 64; blue, 221 }  ,draw opacity=1  , thick]   (135,230) -- (233,145) ;
\draw [color={rgb, 255:red, 215; green, 0; blue, 21 }  ,draw opacity=1  ,thick ]   (135,95) -- (360,185) ;
\draw [dash pattern={on 4.5pt off 4.5pt}, color={rgb, 255:red, 0; green, 64; blue, 221 }  ,draw opacity=1  , thick] (233,236) -- (233,145) ;

\draw (340,240) node [anchor=north west][inner sep=0.75pt]    {$\log \sigma $};
\draw (215,240) node [anchor=north west][inner sep=0.75pt] [color={rgb, 255:red, 0; green, 64; blue, 221 }, opacity=1 ] {$\log \sigma_K $};
\draw (75,75) node [anchor=north west][inner sep=0.75pt]    {$\log \Di^\sigma$};
\draw (180,90) node [anchor=north west][inner sep=0.75pt]  [color={rgb, 255:red, 215; green, 0; blue, 21 }  ,opacity=1 ]  {$\varpropto \sigma ^{-1}$};
\draw (150,169) node [anchor=north west][inner sep=0.75pt]  [color={rgb, 255:red, 0; green, 64; blue, 221 }  ,opacity=1 ]  {$\varpropto \sigma ^{2}$};
\draw (280,120) node [anchor=north west][inner sep=0.75pt]  [color={rgb, 255:red, 36; green, 138; blue, 61 }  ,opacity=1 ]  {$\varpropto \sigma ^{0}$};

\end{tikzpicture}
    \caption{Variation (schematic) of the inertial dissipation $\Di^\sigma (\bs{x})$ as a function of the scale $\sigma$ for a discontinuity (red line), turbulent fluctuations (green line), and viscous/resistive damping (blue line). The intersection between the green and the blue lines defines the dissipative (i.e., Kolmogorov) scale and is noted $\sigma_K$. 
    Similarly, the intersection between the green and the red line can define the discontinuity scale below which discontinuities become dominant (see Figure \ref{fig:ShocksMean}).}
    \label{fig:schemaDi}
\end{figure}

Therefore, depending on the scaling that would be measured in the solar wind (see below) it will be possible to make a distinction between turbulence, viscous/resistive damping and discontinuities (see Figure\,\ref{fig:schemaDi}). Note, however, that other $\sigma$-dependence are possible for non-regular fields \citep{Jaffard2006,Lashermes2008,Jaffard2009}. 

To conclude, we point out that $\Di$ is a generalized function (i.e., a distribution) and its analytic form (if it can be found) can lead to the appearance of a $\delta$-function (see e.g. \cite{DG21}). This means that when the limit $\sigma \to 0$ is taken, one expects to see the value of $\left| \Dis \right|$ increases without limit, however, in practice, the value $\sigma = 0$ will never be reached (see below). 

\section{Methods}
\label{sec:methods}
\subsection{Data selection}

In a first step, we used the \textit{THEMIS-B/ARTEMIS P1} spacecraft data during time intervals when it was traveling in the free streaming solar wind.
The magnetic field data and plasma moments (protons density and velocity) were measured respectively by the Flux Gate Magnetometer (FGM) and the Electrostatic Analyzer (ESA).
All data are expressed in the Geocentric Solar Ecliptic (GSE) coordinate system, have a time resolution $dt=3$s, which corresponds to the spacecraft spin period. We analyzed more than 180 hours of data between 2008 and 2011 that cover both fast and slow solar winds.
Fast winds are defined as having an average speed $U_\mathrm{SW}>450$ km s$^{-1}$.
The others are the slow winds.

In a second step, we analyze PSP's data measured between 2018–2020 during the first and fifth approaches of the spacecraft to the Sun. We selected two subsets of a total duration of about 115 hours corresponding roughly to radial distances of 36 and 30 solar radii (at perihelion) to which we refer respectively by subsets PSP1 and PSP5.
The magnetic field and plasma moments (protons density and velocity) were measured respectively by the fluxgate magnetometer (MAG) and the Solar Probe Analyzer (SPAN). 
All data are expressed in the Radial Tangential Normal (RTN) coordinate system, have a time resolution $dt=1$s

\subsection{Data processing}

For both spacecraft, the selected intervals are divided into samples of two hours, which correspond to a number of data points $N=2400$ for THEMIS-B and $N=7200$ for PSP.
The data selection yielded :
\begin{itemize}
    \item[\labelitemii] 51 samples (122,400 data points) in the slow solar wind.
    \item[\labelitemii] 46 samples (110,400  data points) in the fast solar wind.
    \item[\labelitemii] 61 samples (439,200 data points) for PSP1.
    \item[\labelitemii] 55 samples (396,000 data points) for PSP5.
\end{itemize}

Data gaps (rarely present) were interpolated linearly.
For the selected time intervals, we compute the energy cascade rates $\varepsilon$ estimated by PP98 and the inertial dissipation $\Dis$ using respectively equations (\ref{eq:epsilon}) and (\ref{eq:defDi}).
The structure functions of $\bs{u}$ and $\bs{b}$ are calculated for different time lags $\tau \in \left[1,\: 100\right]dt$ to probe the scales of the inertial range.
We use the Taylor hypothesis $\tau=-\xi/U_\mathrm{SW}$ with $U_\mathrm{SW}$ the mean solar wind speed on the interval, assuming that $\Di = \Di^{\sigma_\text{min}}$, with $\sigma_\text{min}$ the minimum accessible value.
We note $\left\langle \Dis \right\rangle$ the time average of the inertial dissipation over the two hours sample.

Mathematically, the inertial dissipation $\Dis$ can be interpreted as a continuous wavelet transform of the third-order structure function $\bs{Y}$ with respect to the wavelet $\varphi$.
The link between the weak formulation and the wavelet transform reveals several advantages of its application to rough turbulent fields.
Indeed, a wavelet transform can be considered as a ``local Fourier transform" and it is suitable for application to inhomogeneous fields.
Thus, it will genuinely deal with the observed breaking of the spatial translation symmetry \citep{Dubrulle19}.
Therefore, we computed $\Dis$ on the entire time interval for 100 values of $\sigma$ as a continuous 1D wavelet transform based on fast Fourier transform -- a \textsc{Matlab} package provided by the toolbox YAWTB \citep{YAWTB}.
The test function $\varphi^\sigma$ is a normalized Gaussian of width $\sigma$, which is convenient because its derivative is exact (more information on the different ways to implement $\Di$ is given in Appendix \ref{sec:appendixA}).
Note that in the implementation of the inertial dissipation, only the terms depending on $\xi$ are computed because the convolution product is performed on this variable and, given the properties of $\varphi^\sigma$, it is obvious that the smoothing of a field independent of $\xi$ leaves the result unchanged.
To minimize the finite window size effects due to the non-periodicity of the data, we artificially extend each time series to twice it size to apply a Gaussian windowing prior to computing its Fourier transform. The final result is obtained in the time domain after an inverse Fourier transform where only the information from the central part of the time series (i.e. the original one of interest) is considered.

\begin{figure*}[t]
    \centering
    \includegraphics[width=0.49\textwidth]{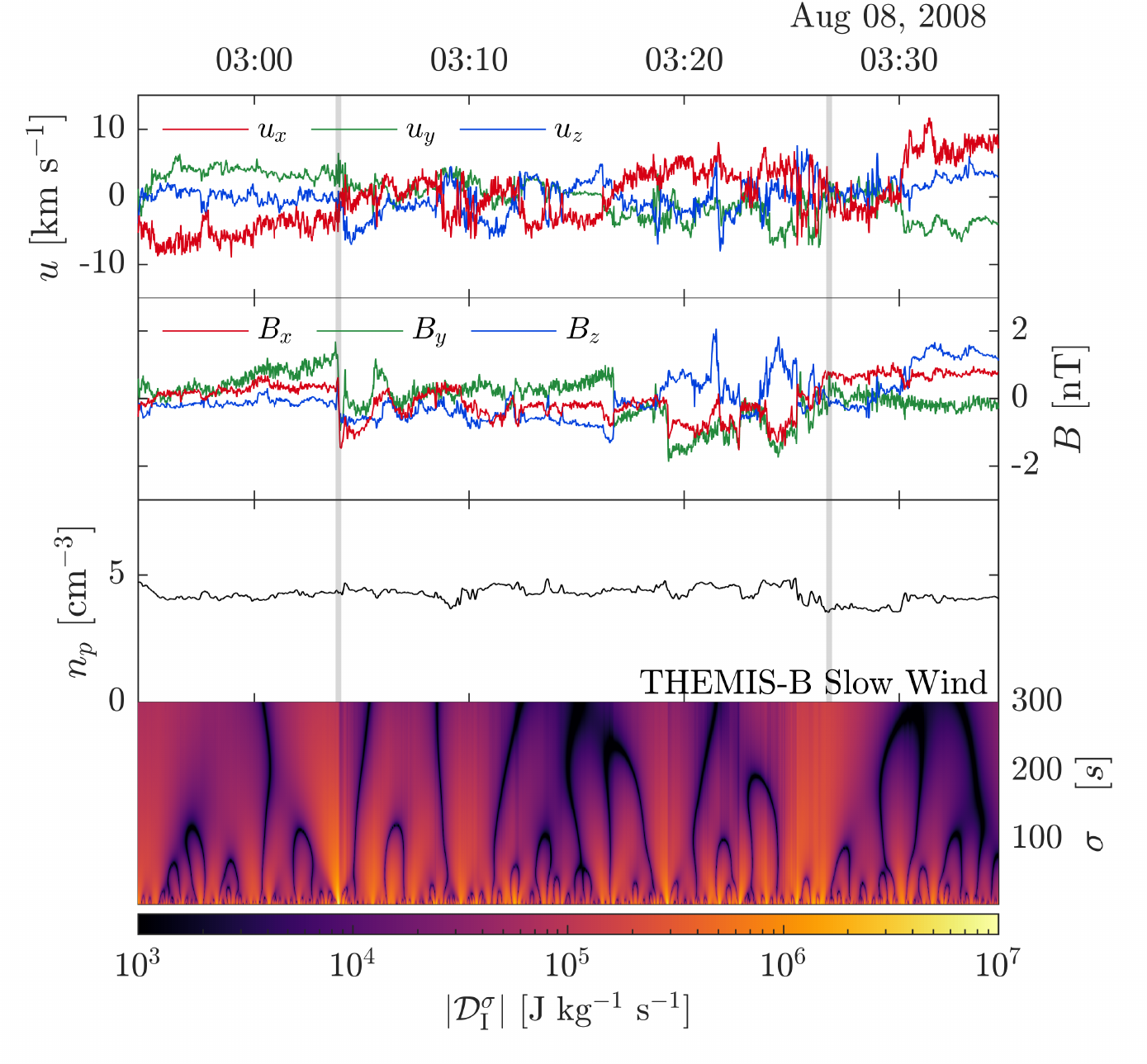}
    \includegraphics[width=0.49\textwidth]{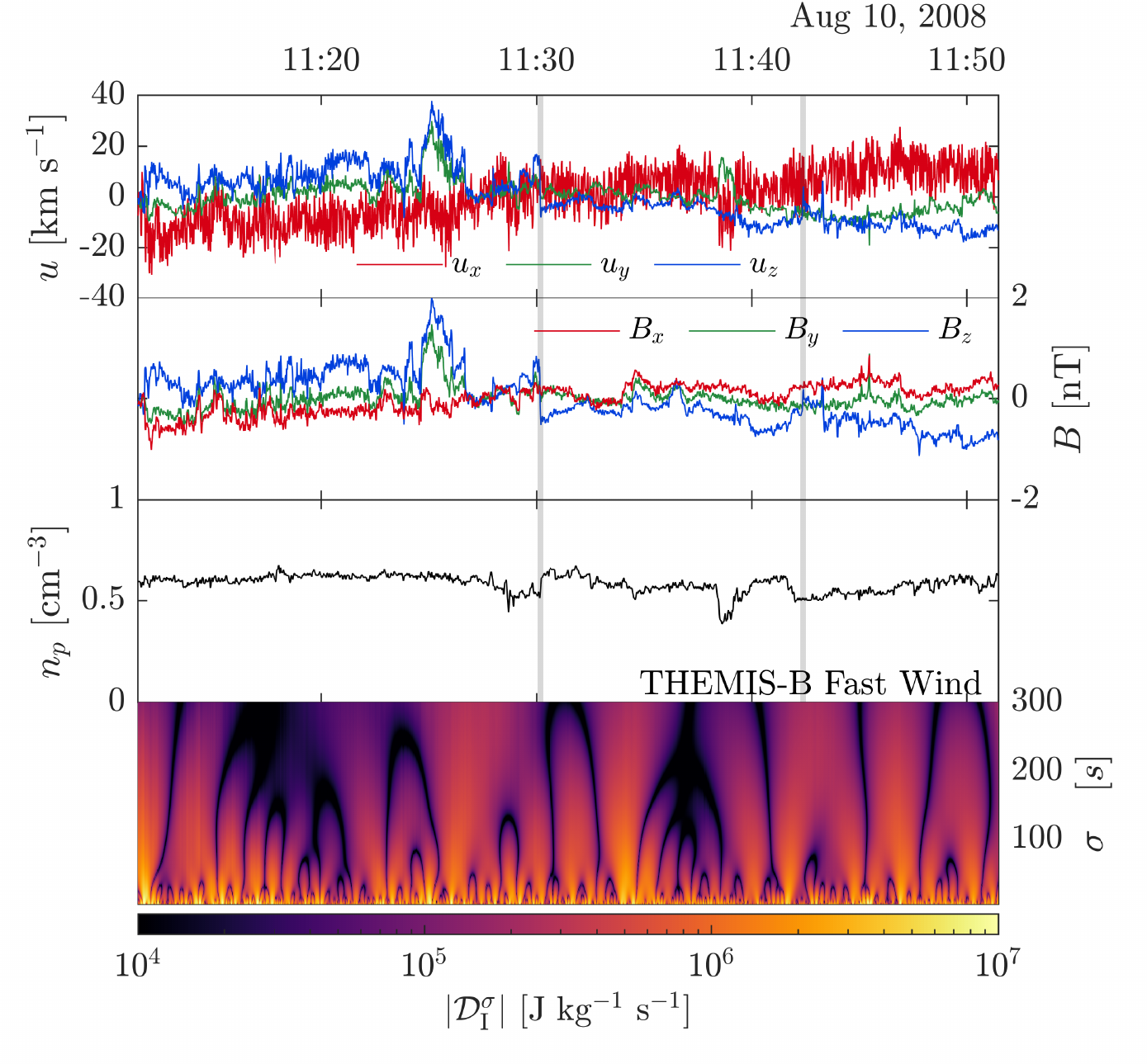}
    \includegraphics[width=0.49\textwidth]{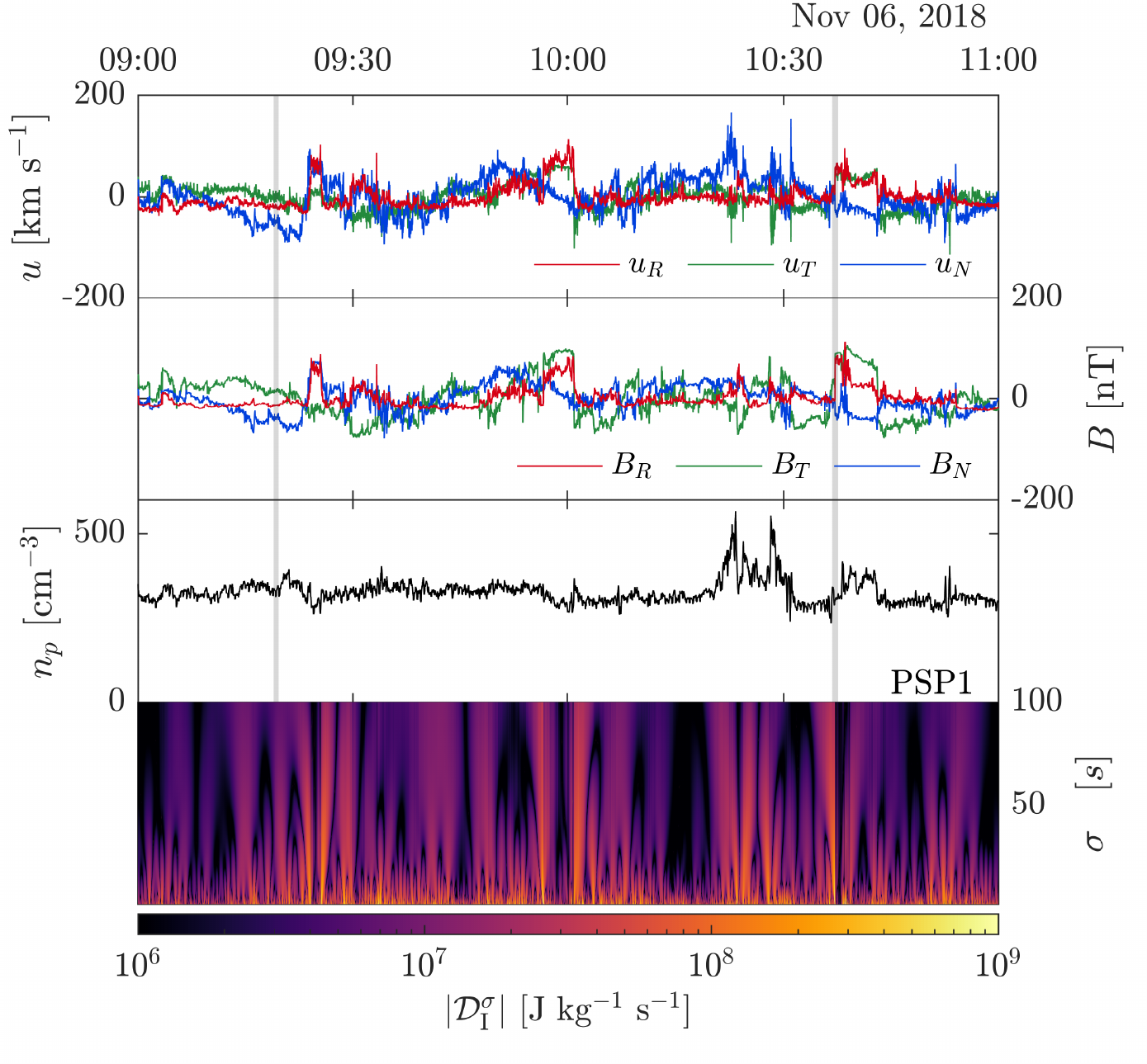}
    \includegraphics[width=0.49\textwidth]{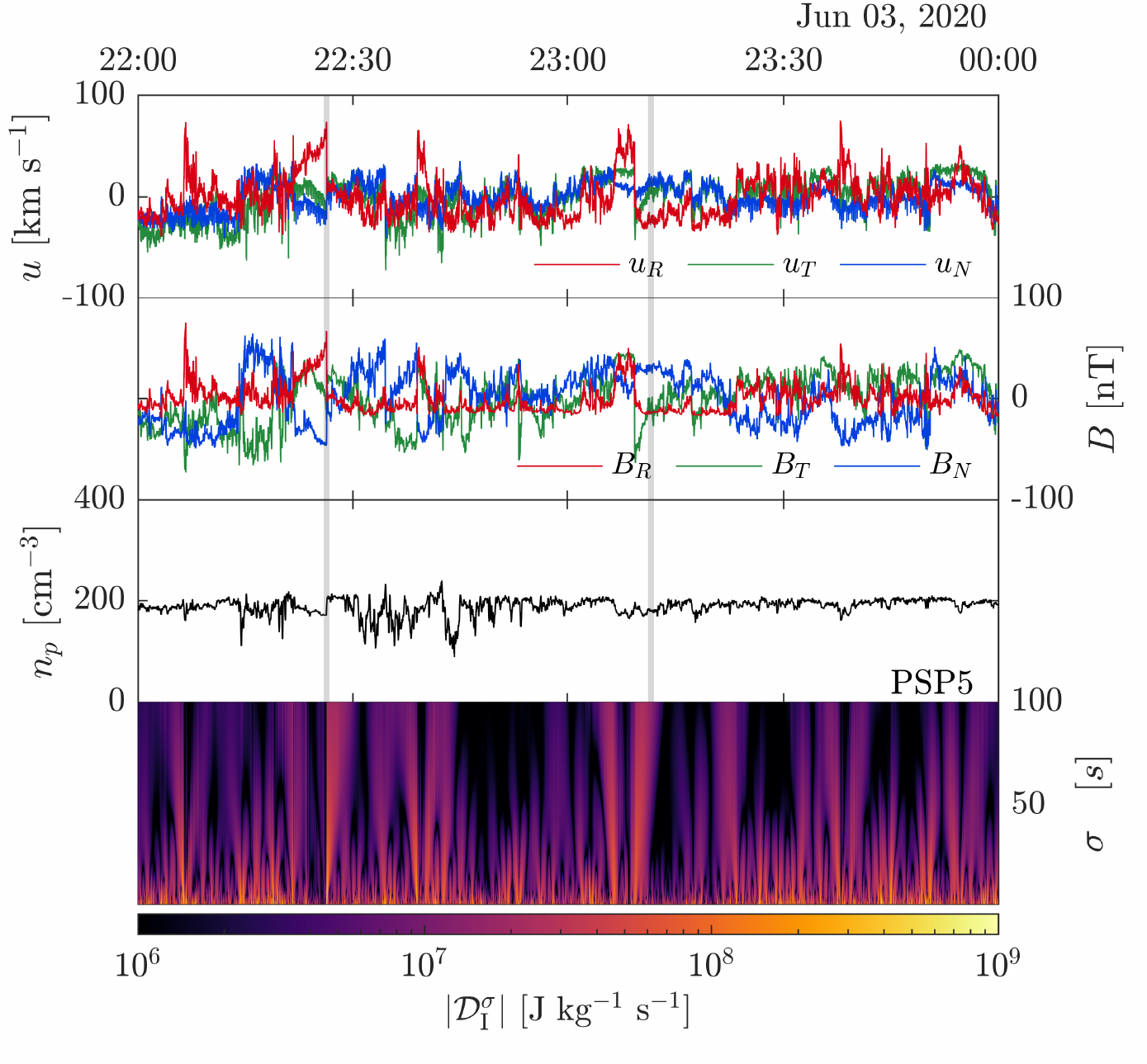}
    \caption{Top panels display the slow (left) and fast (right) winds measured with THEMIS-B. Bottom panels display PSP1 (left) and PSP5 (right). In each panel, from top to bottom, we find the fluctuations of the velocity components, fluctuations of the magnetic field components, proton density and space-scale diagram (in modulus) of the inertial dissipation. The red, blue, and green curves correspond respectively to the $x$, $y$, $z$ components (GSE coordinates) for THEMIS-B and to the $R,T,N$ components (RTN coordinates) for PSP. The vertical gray lines locate the instant for which $\left\vert \Di \right\vert$ is extremal on the sample.}
    \label{fig:ShocksData}
\end{figure*}


\section{Observational results}
\label{sec:results}
\subsection{Inhomogeneous structures}
\label{subsec:inhomogeneousStructures}

We begin our data analysis with four examples where discontinuities are clearly present. In Figure\,\ref{fig:ShocksData} we show
(top left) a THEMIS-B slow wind interval on August 08, 2008 from 02:54:36 to 04:54:36, (top right) a THEMIS-B fast wind interval on April 04, 2011 from 21:15:23 to 23:15:23, (bottom left) a PSP1 interval on November 06, 2018 from 09:00:00 to 11:00:00, and (bottom right) a PSP5 interval on June 03, 2020 from 22:00:00 to June 04, 00:00:00. For each case study, the first two panels (top to bottom) show the three components of the protons velocity and the magnetic field, respectively. They highlight the presence of discontinuities, and thus the breaking of statistical homogeneity, which may jeopardize the use of exact laws.
We find that for the PSP intervals that are closer to the Sun, the velocity and magnetic field components are strongly correlated (respectively 91\%, 90\% and 91\% for the radial, tangential and normal components for the PSP1 interval, and 96\%, 86\% and 80\% for the PSP5 one), which can be interpreted as the signature of outward propagating Alfv\'en waves \citep{Belcher1971}. The third panel shows the proton density, which is relatively constant, and the last panel shows a space-scale diagram of the inertial dissipation (in modulus): time is on the $x$-axis, the width $\sigma$ of the test function on the $y$-axis and the intensity of $\vert \Dis \vert$ is in color.
These maps illustrate the local energy transfer between different scales $\sigma$ (at a given time $t$, or using the Taylor hypothesis, at a given position $x=-U_{SW}t$ with $U_{SW}$ the solar wind speed). If we follow the evolution of the plasma from small to large scales, 
the dark areas delimit the impact of an event on the energy transfer: the larger is the bright area in scale, the greater is the impact of the event in scale and the smaller would be the local energy transfer. Conversely, when a region is mainly dark this means that the energy transfer is local and the dynamics is driven by turbulent fluctuations. 

\begin{figure}
    \centering
    \includegraphics[width=\columnwidth]{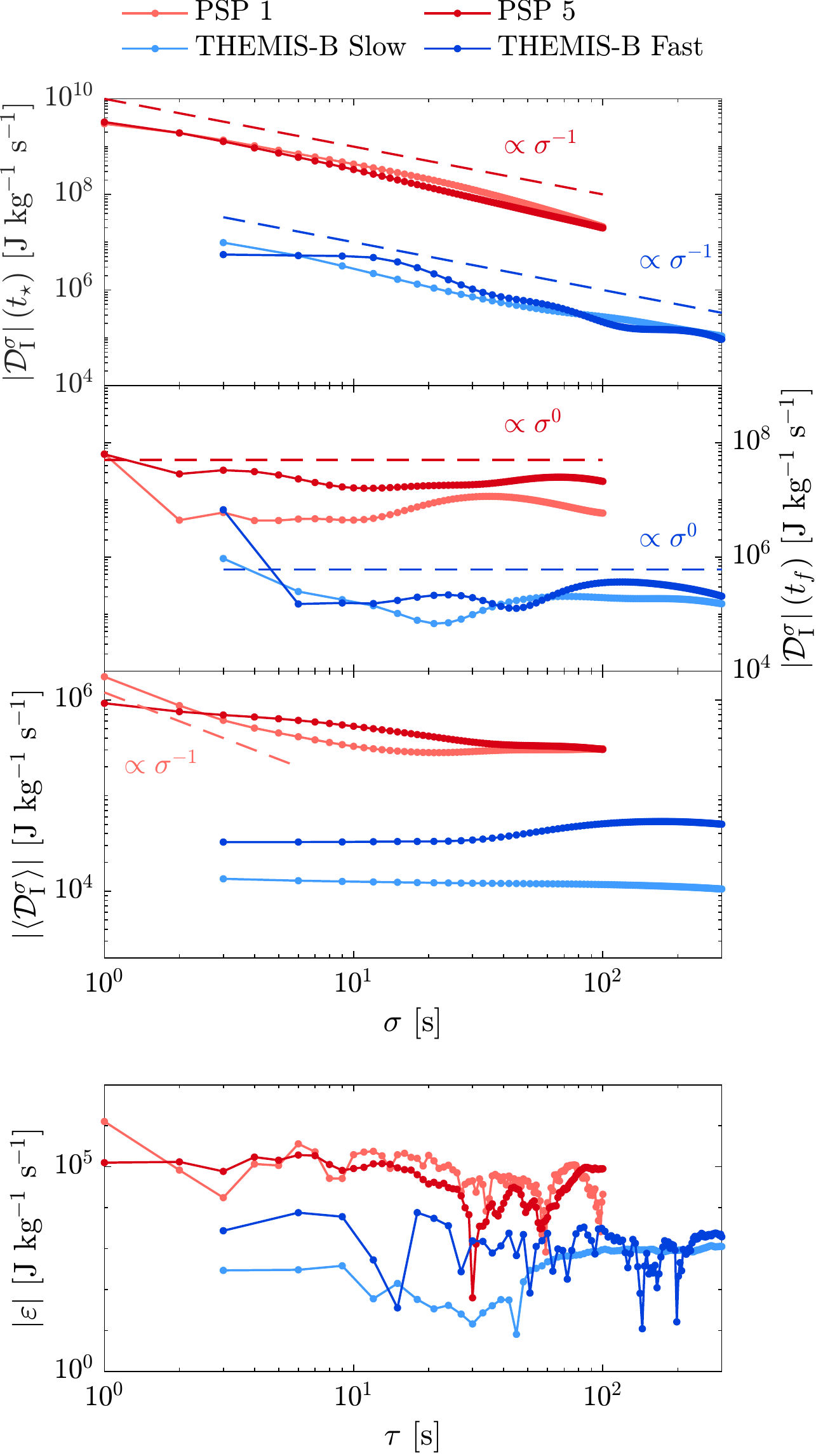}
    \caption{From top to bottom: modulus of the inertial dissipation at time $t_\star$ as a function of scale $\sigma$, modulus of the inertial dissipation at time $t_f$ as a function of scale $\sigma$, estimates of the mean inertial dissipation as a function of $\sigma$, and modulus of the mean rate of energy cascade as a function of $\tau$. Here, $\sigma$ and $\tau$ vary approximately on the same interval.}
    \label{fig:ShocksMean}
\end{figure}

A more precise analysis can be made by observing how $\vert \Dis \vert$ evolves according to the scale $\sigma$ at given times $t_\star$ and $t_f$.
We respectively chose $t_\star$ and $t_f$ such that $\vert \Di(t_\star) \vert = \max \left( \left\vert \Di \right\vert \right)$ and $\vert \Dis(t_f) \vert = \min \left( \left\vert \Di \right\vert \right)$ over the 2h interval (see Figure\,\ref{fig:ShocksData}).
The first and second panels of Figure\,\ref{fig:ShocksMean} reveals that, when placed respectively on a discontinuity (at time $t_\star$) and on a turbulent fluctuation (at time $t_f$), the inertial dissipation does follow the $\sigma^{-1}$ and $\sigma^{0}$ power-laws, as theoretically expected.
The third panel shows the evolution of the inertial dissipation $\vert \meanDis \vert$, averaged over the entire intervals of 2h, as a function of $\sigma$.
The power-laws found indicate the dominant type of energy transfer.
For those coming from THEMIS-B (in blue), we observe mainly a flat profile which means that the dominant mechanism is a turbulent cascade due to fluctuations. For PSP1 (light red), a power law in $\sigma^{-1}$ appears at small $\sigma$, showing the prevalence of discontinuities at small scales for this interval. For PSP5 (dark red), an intermediate power law is observed suggesting that the effect of discontinuities is weaker. 
The bottom panel displays the value of $\vert \varepsilon \vert$ as a function of $\tau$ for the four intervals.
We can see that the curves do not exhibit a clear plateau as theoretically expected; this might be due to the violation of one (or more) of the assumptions on which the exact law formalism is grounded. This is particularly the case for the statistical homogeneity which is unlikely to be valid here because of the presence of discontinuities that distort the estimate of the mean rate of energy cascade \citep{Hadid17}.
Note that for the PSP intervals close to the Sun, both intervals give the same order of magnitude of the inertial dissipation, but is larger than that from Themis data at 1 AU, which overall remain true for the other intervals.
This is consistent with the the radial increase of the turbulent cascade rate $\varepsilon$ as one approaches the Sun \citep{Andres2021,Bandyopadhyay2020a}.
Also the inertial dissipation is larger for fast than for slow solar winds in agreement with previous results regarding the cascade rate $\varepsilon$ \citep{Hadid17}.

\subsection{Switchbacks}
\label{subsec:Switcbacks}

\begin{figure}
    \centering
    \includegraphics[width=\columnwidth]{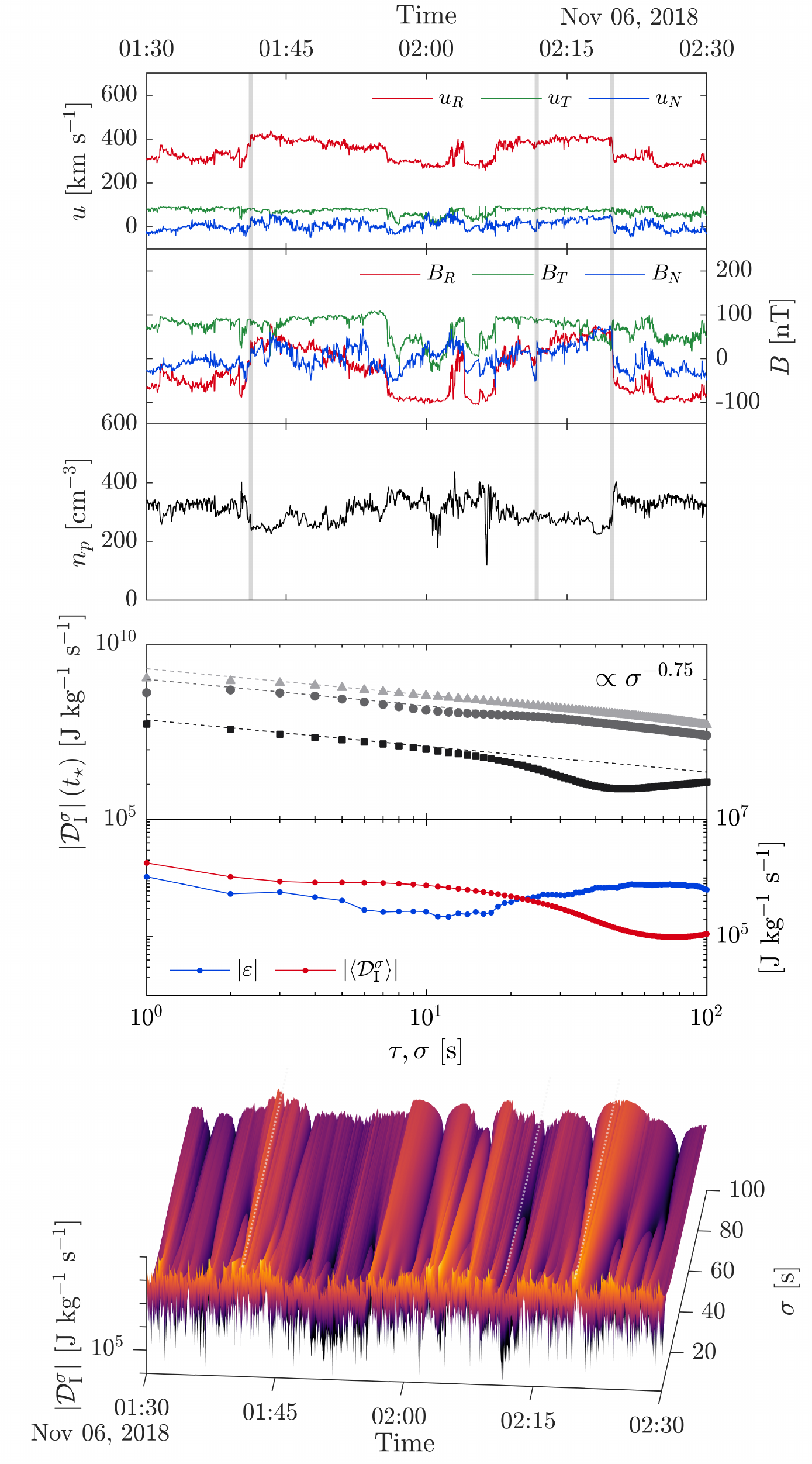}
    \caption{$1$h interval of PSP1 with switchbacks. From top to bottom: velocity components, magnetic field components, proton density, modulus of inertial dissipation (at different times (see also the vertical grey lines in the first three panels and dotted white lines in the last one) $t_\star=$ \{01:41:13, 02:11:47, 02:19:53\} in grey, black and light grey, respectively) as a function of $\sigma$, modulus of $1$h-averaged inertial dissipation as a function of $\sigma$ (red) and modulus of mean rate of energy cascade as a function of $\tau$ (blue), and finally the 3D map of the modulus of inertial dissipation where the color is related to the intensity and thus to the height of $\left\vert \Dis \right\vert$. Velocity and magnetic fields are expressed in RTN coordinates.}
    \label{fig:psp1Switchback}
\end{figure}

Switchbacks are defined as sudden reversals of the radial magnetic field component associated with sharp variations in the radial plasma flow \citep{Neugebauer13, Horbury18,Horbury20}. Although they are actively studied, their origin remains an open question \citep{Bale19,Squire20}. We propose here to estimate the inertial dissipation produced by these peculiar structures in order to quantify their relative importance in the energy cascade.

We focus on a PSP1 interval on November 06, 2018 from 01:30 to 02:30 where switchbacks are numerous.
The first two panels of Figure\,\ref{fig:psp1Switchback} again highlight a clear correlation between the velocity and the magnetic field (respectively 97\%, 86\% and 90\% for the radial, tangential and normal components), which testifies to the presence of outward Alfv\'en waves.
By following the evolution of $\vert \Dis \vert$ as a function of $\sigma$ on switchbacks located at times $t_\star$, a power-law close to $\sigma^{-3/4}$ seems to  emerge. This does not correspond to any scaling laws presented in Section \ref{sec:theoreticalModel} and is therefore not described theoretically by the third-order structure function. The fifth panel shows mainly a flat curve for both the mean rate of energy cascade and the inertial dissipation. We also see that the values coincide relatively well in the limit of small scale $\sigma$. 
The fact that $\varepsilon$ is relatively smooth and constant may come from the fact that the discontinuities are so large that they impose at all scales their jump (or amplitude) on the increments $\delta \bs{u}$ and $\delta \bs{b}$, which then would lead to a higher value of $\varepsilon$ (compared to Figure\,\ref{fig:ShocksMean}).
Although both estimates ($\vert \meanDis \vert$ and $\vert \varepsilon \vert$) give a similar result, rigorously speaking, the exact law should not be applicable in this type of data.
The last panel is a 3D space-scale diagram of inertial dissipation which highlights that switchbacks make the main contribution to the energy cascade. Indeed, one can observe that the large-scale contribution of the inertial dissipation comes from the locations where switchbacks occur and, we observe the same behavior as in subsection \ref{subsec:inhomogeneousStructures}: the dark areas mark the limit of the impact of a discontinuity on its vicinity. 
Overall, we observe that the values of $\vert \Dis \vert$ for switchbacks -- in particular in the limit of small $\sigma$ -- are significantly higher than the values found for the other types of singularities (caracterized by other power-laws -- see also the end of Section \ref{section22}), which suggests that switchbacks can contribute to a stronger heating.

\subsection{Statistical results}
\label{subsec:StatisticalResults}

We conclude our data analysis with a statistical comparison between the mean inertial dissipation and the mean rate of energy transfer as a function of the solar wind speed and the level of the magnetic field fluctuations. Note that the latter is estimated by the ratio between the root mean square $B_{\rm{RMS}}$ and the mean value $B_0$ of the magnetic field. 

\begin{figure}[t]
    \centering
    \includegraphics[width=\columnwidth]{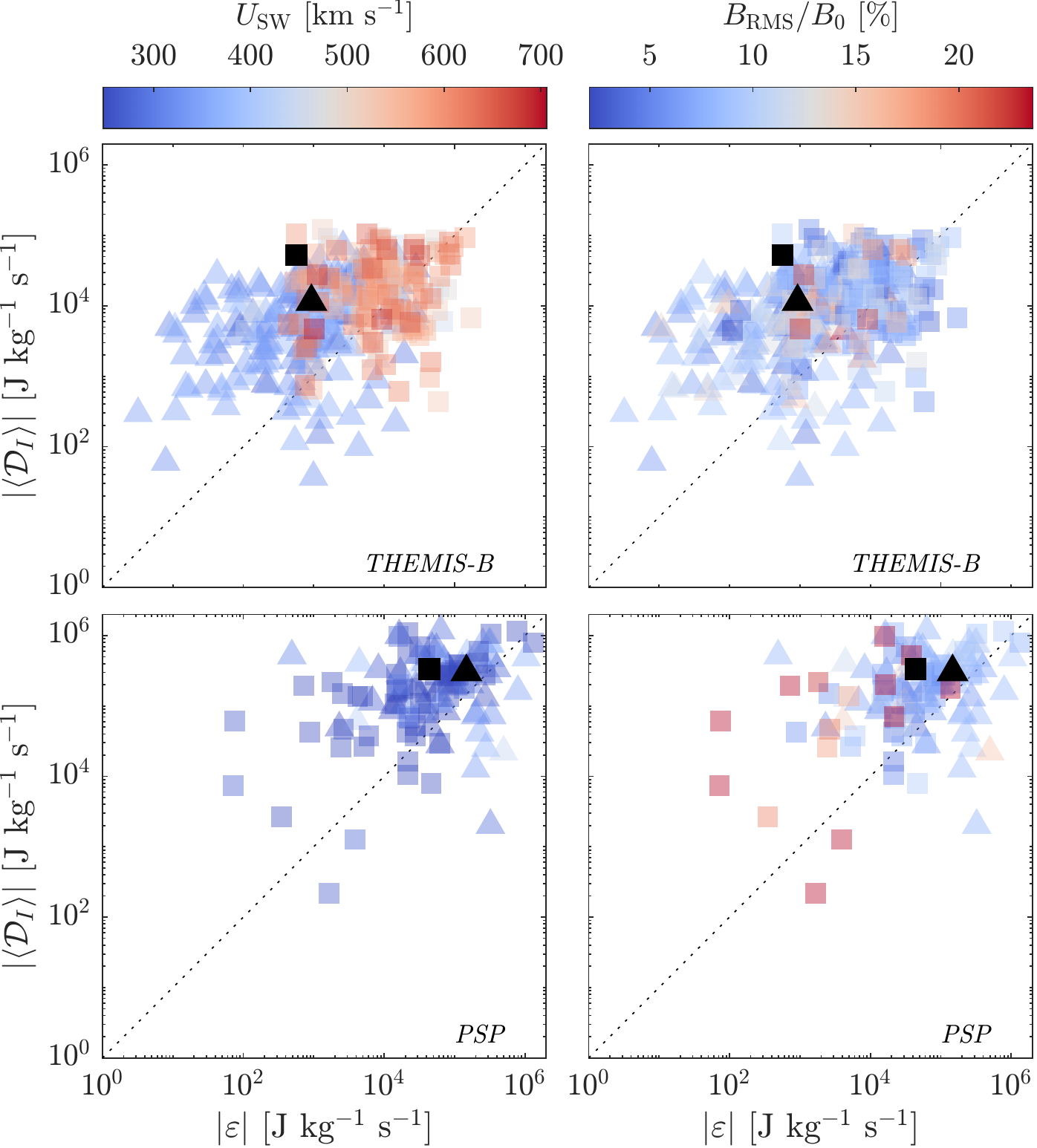}
    \caption{Inertial dissipation as a function of the mean rate of energy transfer measured via the PP98 law.
    The colour scales correspond to the solar wind velocity (left) and to the magnetic field fluctuations (right).
    The triangle and square markers respectively refer to the slow and fast winds (THEMIS-B) in the upper panels, and to PSP1 and PSP5 in the lower panels.
    The dashed (diagonal) lines correspond to $\left\vert \meanDi \right\vert = \left\vert \varepsilon \right\vert$ and black markers are the intervals studied in Figure\,\ref{fig:ShocksData}.}
    \label{fig:heating}
\end{figure}

In Figure \ref{fig:heating}, we show $\vert \meanDi \vert$ as a function of $\vert \varepsilon \vert$ for each processed interval. The upper panels correspond to THEMIS-B intervals (triangles for slow wind and squares for fast wind) and the lower panels to PSP intervals (triangles for PSP1 and squares for PSP5). The dashed (diagonal) line obeys the equation $\left\vert \meanDi \right\vert = \left\vert \varepsilon \right\vert$.
The colors in the left column reflect the mean solar wind velocity while those in the right column correspond to the amplitude of the magnetic field fluctuations of each of the intervals.
First, we notice that near the Sun (bottom panels), the values of $\left\vert \meanDi \right\vert$ and of $\left\vert \varepsilon \right\vert$ are higher than near the Earth (top panels). This property can be attributed primarily to the strength of magnetic field which intensifies as one approaches the Sun, but also to the omnipresence of discontinuities near the Sun.
Note that the decrease of the cascade rate with the heliocentric radial distance has already been measured from exact laws or turbulence transport models, but it is believed that we can only reach a qualitative answer with these models in the presence of discontinuities.
Second, a clear correlation with the wind speed is found at 1\,AU with the two methods: the faster the wind, the higher the mean rate of energy transfer. This property was also shown by \cite{Hadid17} using exact (compressible and incompressible) laws. Note that only THEMIS-B data include fast winds (PSP orbits near the Sun remain mainly in the equatorial plane where the wind is generally slow). 
Third, in the right column, no clear behavior emerges in the magnetic field fluctuations at 1\,AU while for the PSP intervals, even if these events are a few and thus statistically meaningless, large values of $B_\mathrm{RMS}/B_0$ tend to reduce the mean rate of energy transfer (see also Figure \ref{fig:appendixB} in Appendix B).
Last, the majority of the values lies above the diagonal, meaning that on average $\left\vert \meanDi \right\vert > \left\vert \varepsilon \right\vert$. 
This observation can be seen as a signature of inhomogeneities (discontinuities) that are not well captured by the method using the exact law. These inhomogeneities lead mainly to a non-local contribution visible at large $\sigma$ (see Figures  \ref{fig:ShocksData} and \ref{fig:psp1Switchback}).


\section{Discussion \& Conclusion}
\label{sec:conclusion}

In this paper, we have used two different methods (or exact laws) to measure the rate of turbulent energy transfer at MHD scales. The first is the PP98 exact law applicable to homogeneous turbulence, and the second is the local inertial dissipation $\Dis$. Both laws have a similar form with the same combination of structure functions, but in the latter case the homogeneity assumption is not necessary for its derivation. Therefore, $\Dis$ can be considered as more general than the PP98 law since it is a local (exact) law allowing us to measure the energy transfer rate at each point of the turbulent flow even when discontinuities are present. 
Note that the weak formulation of the PP98 law provides a theoretical justification of the observational work of \cite{SorrisoValvo18,SorrisoValvo19a,SorrisoValvo19b}.

Theoretically, several scaling behaviors are expected for $\Dis$ depending to the type of signals. For pure turbulent fluctuations for which the PP98 applies well, a flat signal is expected for $\Dis$ and reported in our study. In the presence of discontinuities, a scaling in $\sigma^{-1}$ is expected and indeed well observed over the whole available range of scales. However, no signature of a dissipation range in $\sigma^{2}$ is detected. These properties can be explained by the fact that the present study is limited to MHD scales. Therefore, a natural extension of this work would be to study sub-MHD scales using data that have the required high time resolution, such as those of the MMS mission, to see if a  $\sigma^{2}$ dissipation can be detected. Unlike the viscous dissipation discussed in Section 2, in collisionless plasma the dissipation involves a complex physics at kinetic scales and a variation different from $\sigma^{2}$ (but still with a positive slope) is likely. The method based on inertial dissipation can offer an original diagnosis to characterize this dissipation. 

Inertial dissipation has many advantages over the exact law but its implementation on real data calls for some caution. This is because the dissipation formula is derived in the theoretical limit $\sigma \to 0$, which is unattainable in real data. The smallest scale that can be used in spacecraft (or simulations) data is set by the available time (or grid) resolution. To what extent the inertial dissipation estimated at this smallest {\it accessible} scale is representative of dissipation at the {\it actual} smallest scale of the system remains thus subject to caution. 

The other limitation of the present study is that it is based on the MHD model. However, this limitation can (partly) be overcome by using the incompressible Hall-MHD model already derived by \cite{G18}, which would allow to probe finer scales and to possibly highlight a correlation between the inertial dissipation with temperature, or to estimate the importance of the Hall effect in the energy cascade.  A further potential improvement is to account for density fluctuations and see how they would impact the inertial dissipation estimates in the solar wind. Such a model remains yet to be derived. However, even with such general models, there will always be a limitation imposed by the temporal resolution of the data that will prevent the strict application of $\sigma \to 0$.

A final caveat that should be kept in mind when estimating both the inertial dissipation and the cascade rate from the exact law, which is inherent to the use of single spacecraft data, is the validity of the Taylor hypothesis and, {\it even when it is valid}, how its use would impact the measured quantities. In the case of the inertial dissipation, the use of the Taylor hypothesis implies that $\Di$ only depends on one dimensional space variable. One can assume isotropy (as done in exact law studies) but this assumption is poorly verified in the solar wind.

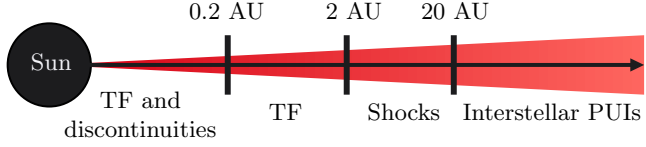
\begin{figure}[h!]
    \centering   

\tikzset {_r3h0dznl2/.code = {\pgfsetadditionalshadetransform{ \pgftransformshift{\pgfpoint{0 bp } { 0 bp }  }  \pgftransformrotate{0 }  \pgftransformscale{2 }  }}}
\pgfdeclarehorizontalshading{_nsk9tg5c1}{150bp}{rgb(0bp)=(0.84,0,0.08);
rgb(37.5bp)=(0.84,0,0.08);
rgb(37.5bp)=(0.84,0,0.08);
rgb(62.5bp)=(1,0.41,0.38);
rgb(100bp)=(1,0.41,0.38)}
\tikzset{every picture/.style={line width=0.75pt}} 

\begin{tikzpicture}[x=0.75pt,y=0.75pt,yscale=-0.6,xscale=0.6]

\path  [shading=_nsk9tg5c1,_r3h0dznl2] (100,155) -- (600,130) -- (600,180) -- cycle ; 
 \draw  [color={rgb, 255:red, 215; green, 0; blue, 21 }  ,draw opacity=0 ] (100,155) -- (600,130) -- (600,180) -- cycle ; 

\draw  [fill={rgb, 255:red, 28; green, 28; blue, 30 }  ,fill opacity=1 ] (65,155) .. controls (65,135.67) and (80.67,120) .. (100,120) .. controls (119.33,120) and (135,135.67) .. (135,155) .. controls (135,174.33) and (119.33,190) .. (100,190) .. controls (80.67,190) and (65,174.33) .. (65,155) -- cycle ;
\draw [color={rgb, 255:red, 28; green, 28; blue, 30 }  ,draw opacity=1 ][line width=1.5]    (100,155) -- (596,155) ;
\draw [shift={(600,155)}, rotate = 180] [fill={rgb, 255:red, 28; green, 28; blue, 30 }  ,fill opacity=1 ][line width=0.08]  [draw opacity=0] (11.61,-5.58) -- (0,0) -- (11.61,5.58) -- cycle    ;
\draw [color={rgb, 255:red, 28; green, 28; blue, 30 }  ,draw opacity=1 ][line width=2.25]    (440,180) -- (440,130) ;
\draw [color={rgb, 255:red, 28; green, 28; blue, 30 }  ,draw opacity=1 ][line width=2.25]    (350,180) -- (350,130) ;
\draw [color={rgb, 255:red, 28; green, 28; blue, 30 }  ,draw opacity=1 ][line width=2.25]    (250,180) -- (250,130) ;

\draw (82,143) node [anchor=north west][inner sep=0.75pt]   [align=left] {\textcolor[rgb]{0.95,0.95,0.97}{Sun}};
\draw (110,175) node [anchor=north west][inner sep=0.75pt]   [align=center] {TF and \\ discontinuities};
\draw (282,185) node [anchor=north west][inner sep=0.75pt]   [align=left] {TF};
\draw (365,185) node [anchor=north west][inner sep=0.75pt]   [align=left] {Shocks};
\draw (445,185) node [anchor=north west][inner sep=0.75pt]   [align=left] {Interstellar PUIs};
\draw (215,100) node [anchor=north west][inner sep=0.75pt]   [align=left] {0.2 AU};
\draw (330,100) node [anchor=north west][inner sep=0.75pt]   [align=left] {2 AU};
\draw (410,100) node [anchor=north west][inner sep=0.75pt]   [align=left] {20 AU};

\end{tikzpicture}
    \caption{Schematically, the heliospheric turbulence can be separated into four regions where the mean rate of energy transfer has different origins. TF and PUIs stand for turbulent fluctuations and pickup ions, respectively. Note that this classification is made in terms of variations in the basic fields that enter the MHD equations. Therefore, this view is more rooted in the physics of turbulence than in the sources of turbulence of the solar wind.}
    \label{fig:schemaSunHeating}
\end{figure}

Several heating mechanisms exist in the solar wind (see Figure \ref{fig:schemaSunHeating}) and their predominance seems to depend on the heliospheric radial distance as shown by the proton temperature measurements (with a slow decrease of the temperature up to 20 AU, then an increase beyond 20 AU \citep{Matthaeus99,Elliott2019}). It is well known that around 1\,AU turbulent fluctuations are dominant, but closer to the Sun both discontinuities and strong turbulent fluctuations are important as now evidenced in PSP observations, while beyond 2\,AU we observe large-scale inhomogeneous structures such as interplanetary shocks, with relatively weak turbulent fluctuations. 
Beyond 20\,AU, the dominant heating mechanism is mainly pickup ions \citep{Zank2018,Pine20b}.
Faced with such a variety of processes, it is interesting to have a tool that allows us to quantify the turbulent energy cascade rate at fluid scales, regardless of the dominant heating mechanism at work. The inertial dissipation seems to be a good candidate for this purpose.

\begin{acknowledgments}
Aknowledgments : V.D. acknowledges B. Dubrulle for helpful discussion. 
\end{acknowledgments}


\appendix

\section{Comparison of algorithms for computation of inertial dissipation}
\label{sec:appendixA}
To compute equation (\ref{eq:defDi}), different possibilities are available.
The first one, and the one chosen for this work, is to apply the gradient on the test function $\varphi^\sigma$.
The latter being known analytically, its implementation does not introduce any numerical error and respect the hypothesis of non-regularity of the fields at the origin of the derivation of $\Di$.
A second possibility is to perform an integration by part so that the gradient acts on the structure function $\bs{Y}$.
The form obtained is almost identical to the PP98 law before integration assuming isotropy but, on the one hand, this is in contradiction with the assumption of non-regularity of the fields and, on the other hand, it introduces numerical errors when computing its gradient.
\begin{figure}[h]
    \centering
    \includegraphics[scale=.75]{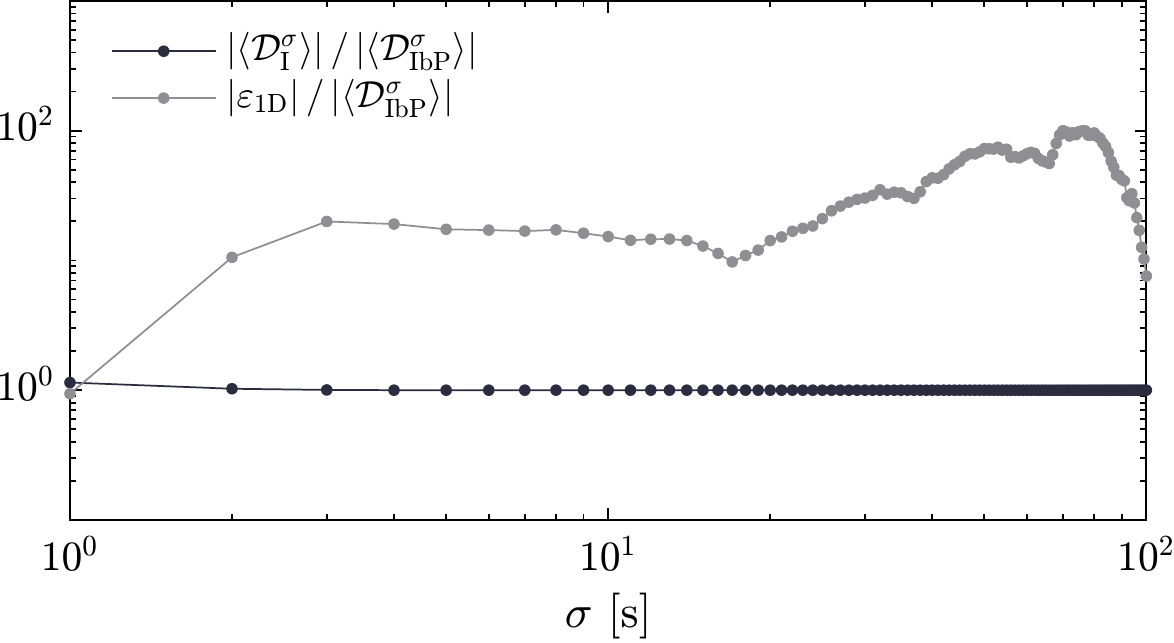}
    \caption{Evolution of $ \left\vert \left\langle \mathcal{D}_\mathrm{I}^\sigma \right\rangle  \right\vert / \left\vert \left\langle \mathcal{D}_\mathrm{IbP}^\sigma \right\rangle \right\vert$ and $ \left\vert \varepsilon_\mathrm{1D}  \right\vert / \left\vert \left\langle \mathcal{D}_\mathrm{IbP}^\sigma \right\rangle \right\vert$ as a function of $\sigma$. For a consistent comparison, the time lag $\tau$ involved in the computation of $\varepsilon_\mathrm{1D}$ takes the same values as $\sigma$.}
    \label{fig:appendixA}
\end{figure}

To verify in practice the difference between these two computations, we compared the estimation of the inertial dissipation with and without integration by parts (hereafter named $\Di$ and $\mathcal{D}_\mathrm{IbP}$ respectively) as well as PP98 without the isotropy assumption, named $\varepsilon_\mathrm{1D}$.
In Figure \ref{fig:appendixA} we show the comparison between these three methods for the interval studied in subsection $\ref{subsec:Switcbacks}$.
The effect of the integration by parts is only slightly felt at small scale because the black curve is equal to 1 for all the values of $\sigma$ except for the minimal one and, the grey curve confirms that when $\sigma \to \sigma_\mathrm{min}$, we find the equality $\left\langle \mathcal{D}_\mathrm{IbP} \right\rangle = \varepsilon_\mathrm{1D}$ predicted theoretically.

\section{Radial evolution of the magnetic field}
\label{sec:appendixB}
\begin{figure}[h]
    \centering
    \includegraphics[width=\textwidth]{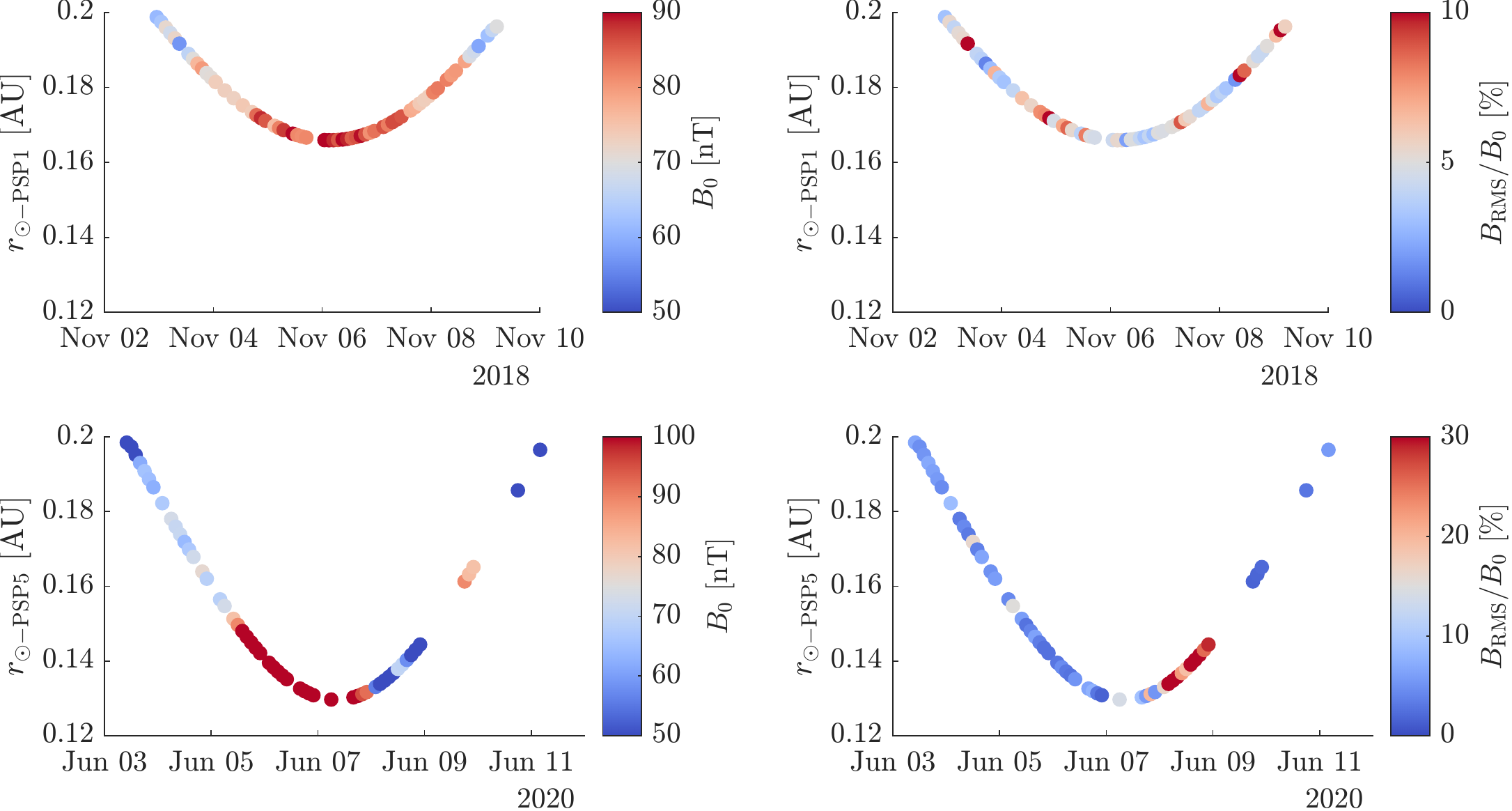}
    \caption{Radial evolution of PSP1 and PSP5 during the first (top) and fifth (bottom) approaches. The color shows the relative intensity of the average value of the magnetic field (left), and its normalized fluctuations (right).}
    \label{fig:appendixB}
\end{figure}

To verify that the lack of correlation underlined in the description of Figure \ref{fig:heating} is not a curiosity, it is interesting to look at the evolution of the magnetic field as PSP approaches the Sun.
Figure \ref{fig:appendixB} shows that as the radial distance decreases, the average magnetic field strength $B_0$ increases and the ratio $B_\mathrm{RMS} / B_0$ decreases.
This is thus consistent with the results of Section 4.

\bibliography{main}{}
\bibliographystyle{aasjournal}

\end{document}